\documentclass[letterpaper,preprint,aps]{revtex4-1}
\usepackage[latin9]{inputenc}
\usepackage{color}
\usepackage{amsmath}
\usepackage{amssymb}
\usepackage{graphicx}
\usepackage[unicode=true,pdfusetitle,
 bookmarks=true,bookmarksnumbered=false,bookmarksopen=false,
 breaklinks=false,pdfborder={0 0 1},backref=false,colorlinks=false]
 {hyperref}

\makeatletter

\pdfpageheight\paperheight
\pdfpagewidth\paperwidth

\usepackage{graphicx}
\usepackage{tikz}
\usepackage{comment}

\makeatother

\begin{document}
\title{BMS Symmetry via AdS/CFT}
\author{David A. Lowe}
\email{lowe@brown.edu}

\affiliation{Department of Physics, Brown University, Providence, RI, 02912, USA}
\altaffiliation{Brown Theoretical Physics Center, Providence, RI, 02912, USA}

\author{David M. Ramirez}
\email{david\_ramirez@brown.edu}

\affiliation{Brown Theoretical Physics Center, Providence, RI, 02912, USA}
\begin{abstract}
With a view to understanding extended-BMS symmetries in the framework
of the $AdS_{4}/CFT_{3}$ correspondence, asymptotically AdS geometries
are constructed with null impulsive shockwaves involving a discontinuity
in superrotation parameters. The holographic dual is proposed to be
a two-dimensional Euclidean defect conformal field localized on a
particular timeslice in a three-dimensional conformal field theory
on de Sitter spacetime. The defect conformal field theory generates
a natural action of the Virasoro algebra. The large radius of curvature
limit $\ell\to\infty$ yields spacetimes with nontrivial extended-BMS
charges. 
\end{abstract}
\maketitle

\section{Introduction}

The extended-BMS symmetry algebra \citep{Bondi:1962px,Sachs:1962wk,Barnich:2009se,Barnich:2010eb,Barnich:2011mi,Strominger:2013jfa,Strominger:2014pwa}
appears to provide an infinite set of conserved quantities in asymptotically
flat spacetime geometries and it is of great interest to understand
to what extent these symmetries constrain the local dynamics and how
these classical quantities generalize to the quantum level. For example,
it has been suggested in \citep{Hawking:2016sgy,Hawking:2016msc,Haco:2018ske}
that such quantities give rise to an infinite amount of quantum hair
on black holes, which would be of profound importance for the information
problem.

The goal of the present work is to study these quantities as a limit
of asymptotically anti-de Sitter spacetimes where one may gain additional
insight via the anti-de Sitter/conformal field theory correspondence.
Some initial works in this direction are \citep{Hijano:2019qmi,Hijano:2020szl}.
It has been shown that the BMS algebra generalizes to a Lie algebroid
in asymptotically anti-de Sitter spacetime \citep{Compere:2019bua,Compere:2020lrt}.
One concludes from this that in general the asymptotically anti-de
Sitter geometries of interest involve nontrivial deformations of the
boundary metric and lead to scenarios which are not well-understood
from the holographic perspective. 

In the present work we take a somewhat different approach to building
a holographic description of asymptotically anti-de Sitter geometries
with the analog of extended-BMS charges. We begin by considering null
spherical impulsive shock waves where the metric across the shock
undergoes a superrotation. The shock reflects off the boundary of
anti-de Sitter spacetime, preserving the standard Dirichlet boundary
conditions on the metric. This induces a nontrivial boundary stress
energy tensor at the point of intersection, which is matched with
the expectation value of the stress energy in the holographic dual.
This suggests the shock should be identified with a two-dimensional
Euclidean defect conformal field theory within a three-dimensional
Lorentzian CFT on the boundary. In this setup, the defect CFT allows
a natural action of Virasoro charges, in addition to the usual action
associated with the global conformal symmetry of the three-dimensional
CFT $SO(3,2$). The central charge of the algebra is non-zero, and
is read off from the solution. 

One may then take a large radius of curvature limit $\ell\to\infty$
of this setup to recover a holographic description of gravity in asymptotically
flat spacetime. The limit is taken so that the defect CFT lives at
spacelike infinity, once conformally compactified. In this limit the
$SO(3,2)$ contracts to $ISO(3,1$) \citep{Compere:2019bua,Compere:2020lrt}.
The defect CFT induces a nontrivial BMS charge when the asymptotically
flat limit is taken. The full symmetry group includes the generators
realized on the defect CFT combined with actions on the fields of
the three-dimensional CFT. The more general supertranslation generators
of the BMS algebra are not so far manifest in this construction.

The shock solution is based on an exact solution which introduces
an analytic diffeomorphism on the sphere \citep{Hogan:1995qb}. These
solutions were also considered in the context of superrotations in
\citep{Strominger:2016wns}. We promote this to an asymptotically
anti-de Sitter solution. The solution has a mild Dirac delta function
curvature singularity in affine null coordinates across the shock,
but is otherwise an exact solution of the vacuum Einstein equations
with negative cosmological constant. 

Much is known about the general structure of conformal field theories,
and there will be a discrete infinity of quasi-primary operators with
positive conformal weights $\Delta$. In the limit $\ell$ to infinity,
the spectrum of the asymptotically flat theory will inherit this discrete
structure. Moreover we obtain a holographic realization of asymptotically
flat spacetimes via a limit of a three-dimensional CFT with a two-dimensional
defect. The HKLL construction \citep{Hamilton:2005ju,Hamilton:2006az,Kabat:2011rz}
should extend straightforwardly to this setup and bulk fields receive
contributions both from smeared primaries in the three-dimensional
CFT as well as operators in the two-dimensional defect CFT, which
alone carry the extended-BMS charges, though we leave the details
of this to future work.

\section{Calculations}

The goal is to gain insight into the flat spacetime limit of holography
by considering the limit where the radius of curvature $\ell\to\infty$
from the Anti-de Sitter side, where the holographic dual is thought
to be a well-defined conformal field theory. Some discussion of the
viewpoint has appeared in \citep{Hijano:2019qmi,Hijano:2020szl}. 

\begin{figure}
\includegraphics[height=12cm]{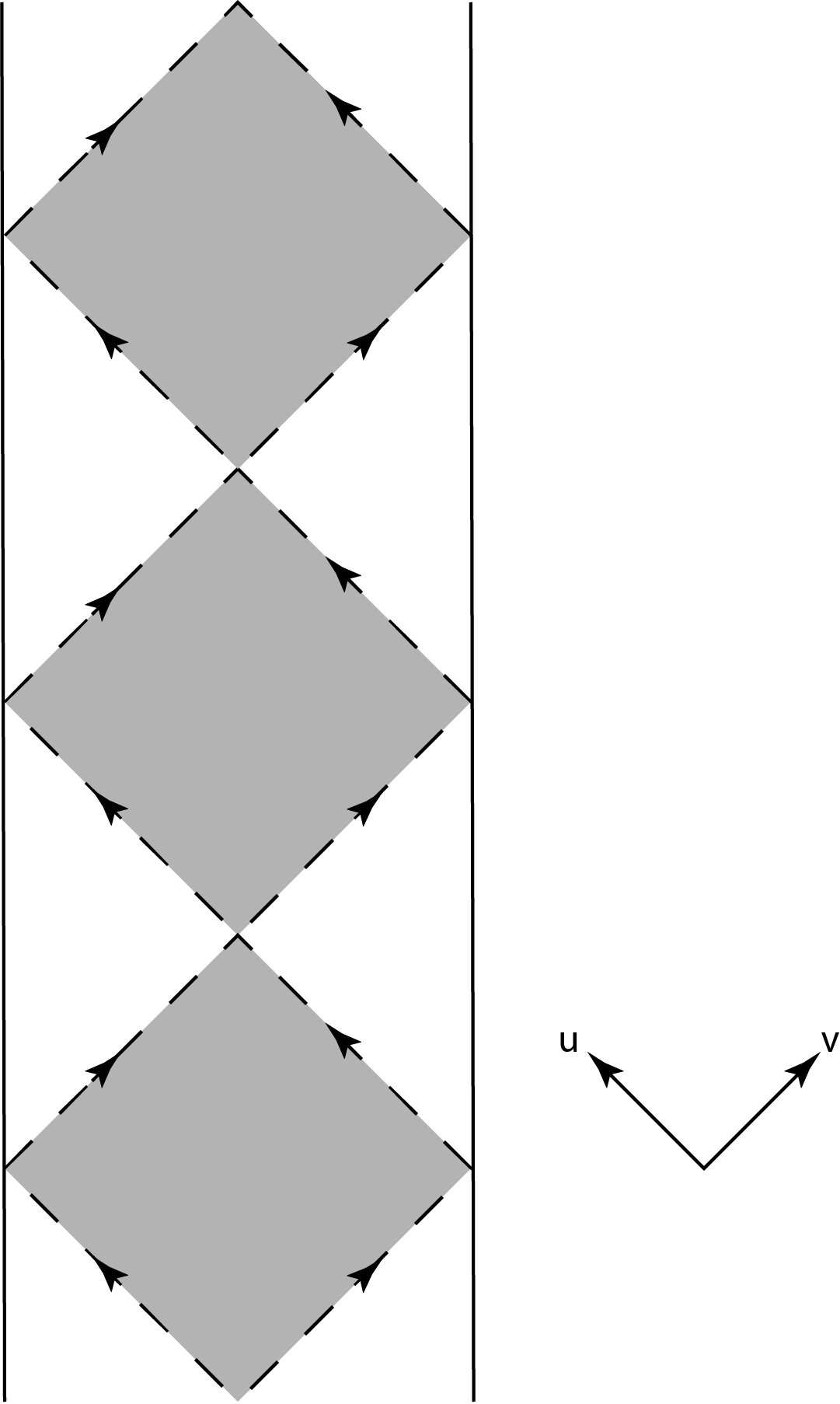}\caption{\label{fig:Penrose-diagram-showing}Penrose diagram showing the asymptotically
anti-de Sitter imploding-exploding impulsive waves. The grey interior
region corresponds to a solution with superrotation charge. The unshaded
regions correspond to the anti-de Sitter vacuum with the unperturbed
metric on the boundary.}

\end{figure}

We will be interested in a set of spherical impulsive gravitational
wave solutions as studied in \citep{Hogan:1992kg,Hogan:1995qb}. These
solutions were also considered in the context of superrotations in
\citep{Strominger:2016wns}. These solutions are most naturally written
in terms of Kruskal-like coordinates. A generalization of the imploding-exploding
solution studied in \citep{Hogan:1995qb} to asymptotically anti-de
Sitter spacetime is\footnote{For convenience we have rescaled the coordinates by factors of $\sqrt{2}$
compared to \citep{Hogan:1995qb}.}
\[
ds^{2}=\frac{\ell^{2}}{(1+uv)^{2}}\left(\frac{4}{\left(1+z\bar{z}\right)^{2}}\left|(u-v)dz+2\left[u\bar{F}(\bar{z})\theta(u)+v\bar{G}(\bar{z})(1-\theta(v))\right]\left(1+z\bar{z}\right)^{2}d\bar{z}\right|^{2}-4dudv\right)
\]
where $\theta(x)$ is the Heaviside step function and $F(z),$ $G(z)$
are Schwarzian derivatives of holomorphic functions $f(z),g(z)$ of
the form
\begin{equation}
F(z)=\frac{f'''(z)}{f'(z)}-\frac{3}{2}\left(\frac{f''(z)}{f'(z)}\right)^{2}\,,\label{eq:schwarzian}
\end{equation}
and similarly for $G(z).$ The holomorphic functions $f(z),g(z)$
parametrize superrotations. This is most easily seen by noting that
across the shock (say at $u=0)$ the metric may be obtained by performing
the coordinate transformation of the ordinary flat spacetime metric
by
\begin{equation}
z\to f(z),\qquad v\to\frac{1+\left|f(z)\right|^{2}}{1+\left|z\right|^{2}}\frac{v}{\left|f'(z)\right|}\,.\label{eq:superrotate}
\end{equation}
For our purposes it will be convenient to transform by a 1/4 period
in the global time coordinate which amounts to a change of coordinates
from \citep{Hogan:1995qb} of
\[
u\to\frac{u+1}{u-1},\qquad v\to\frac{v+1}{v-1}\,.
\]
We also include the reflections of the waves off the boundary of AdS
to yield a solution of the form
\begin{align}
ds^{2} & =\frac{\ell^{2}}{(1+uv)^{2}}\left(\frac{4}{\left(1+z\bar{z}\right)^{2}}\left|(u-v)dz+\left(-(u+1)(1-v)\theta(v-1)\bar{G}+\right.\right.\right.\label{eq:implode_explode}\\
 & \left.\left.\left.(u+1)(1-v)\theta(u+1)\bar{F}\right)\left(1+z\bar{z}\right)^{2}d\bar{z}\right|^{2}-4dudv\right)\,.\nonumber 
\end{align}
The coordinates $(u,v)$ range from $-\infty$ to $\infty$ with $v>u$,
and with the timelike boundary $uv<-1$ and the spacelike coordinate
patch boundary $uv<1$. The figure \ref{fig:Penrose-diagram-showing}
shows a sequence of such solutions glued together along the spacelike
boundaries of the coordinate patches. These impulsive gravitational
waves are solutions of the vacuum Einstein equations with Dirac delta
function curvature singularities along the null lines $u=-1$ and
$v=1$. We allow for stress energy localized on the boundary, which
will be computed below, to enforce the standard Dirichlet boundary
conditions at $uv=-1$. In this case we set $F=G$. 

To become more familiar with these solutions, let us for the moment
consider the case $F=0$ and map to embedding coordinates where (for
simplicity we define our coordinates to be dimensionless, then restore
units by multiplying the metric by $\ell^{2}$ as above)
\[
-T^{2}-X^{2}+R^{2}=-1\,.
\]
Then we can identify
\[
T-R=u(1+X),\qquad T+R=v(1+X)\,.
\]
The metric associated with the 2-sphere is
\begin{equation}
ds_{2}^{2}=R^{2}d\Omega^{2}=\frac{4R^{2}dzd\bar{z}}{\left(1+z\bar{z}\right)^{2}}\label{eq:roundmetric}
\end{equation}
where 
\[
z=e^{i\phi}\tan\frac{\theta}{2}\qquad\bar{z}=e^{-i\phi}\tan\frac{\theta}{2}\,.
\]
Combining these we obtain the metric \eqref{eq:implode_explode} in
the unshaded region. 

The full 3-dimensional boundary at infinity corresponds to the surface
$uv=-1$. The induced metric is then conformal to three-dimensional
de Sitter spacetime with metric
\begin{equation}
ds^{2}=\frac{4}{\left(1+z\bar{z}\right)^{2}}\left|\left(u+\frac{1}{u}\right)dz\right|^{2}-\frac{4}{u^{2}}du^{2}\,.\label{eq:boundarymetric}
\end{equation}
Thus the shock at $u=-1$ meets the boundary at infinity on a spacelike
2-sphere with metric conformal to the standard round metric. It is
interesting to note a similar $dS_{3}$ geometry appears at spacelike
infinity in the construction of the symmetry generators in asymptotically
flat spacetime \citep{Ashtekar:1978zz,Ashtekar_1984,Prabhu:2019daz,Prabhu:2019fsp}.

It will be helpful to re-express the Kruskal coordinates as Fefferman-Graham
coordinates with a 3-dimensional de Sitter boundary metric. To do
this let us define $t$ and $\rho$ 
\[
t=\log\left(-\frac{v}{u}\right)=\log\left(-\frac{T+R}{T-R}\right),\qquad\frac{4\rho}{(1+\rho)^{2}}=1+uv=1+\frac{(T+R)(T-R)}{(1+X)^{2}}
\]
or equivalently
\[
T=\frac{1-\rho^{2}}{2\rho}\sinh\frac{t}{2},\qquad R=\frac{1-\rho^{2}}{2\rho}\cosh\frac{t}{2},\qquad X=\frac{1+\rho^{2}}{2\rho}
\]
then the anti-de Sitter metric becomes
\[
ds^{2}=\ell^{2}\frac{d\rho^{2}+\frac{1}{16}\left(1-\rho^{2}\right)^{2}\left(\left(e^{t/2}+e^{-t/2}\right)^{2}d\Omega^{2}-dt^{2}\right)}{\rho^{2}}\,.
\]
This matches the induced metric on the boundary \eqref{eq:boundarymetric}
at $\rho\to0$ with the change of variables $u=-e^{-t/2}$.

\section{Holographic Stress Tensor}

In this section we study the holographic mapping of the metric near
the boundary of AdS to the stress energy tensor of the CFT. Following
the procedure of \citep{deHaro:2000vlm} the $g_{ab}^{(3)}$ component
of the metric in Fefferman-Graham coordinates
\[
ds^{2}=\frac{\ell^{2}d\rho^{2}+\left(g_{ab}^{(0)}+\rho^{2}g_{ab}^{(2)}+\rho^{3}g_{ab}^{(3)}+\cdots\right)dx^{a}dx^{b}}{\rho^{2}}
\]
is identified with the expectation value of the CFT stress tensor.
Here $x^{a}$ are the transverse coordinates, and $g_{ab}^{(0)}$
is the boundary metric. In this expansion $g_{ab}^{(2)}$ is determined
in terms of $g_{ab}^{(0)}$, but $g_{ab}^{(3)}$ is independent data. 

In the present situation the metric involves step functions, and leads
to a curvature tensor that must be treated as a generalized function.
This takes us beyond the original considerations of \citep{deHaro:2000vlm},
however the main part of the derivation will carry over. To proceed,
we continue to match $g_{ab}^{(0)}$ with the unperturbed metric \eqref{eq:boundarymetric}.
One approach would be to construct the Brown-York tensor \citep{Brown:1992br}
with the usual counter-terms at $\rho=\epsilon$. However this surface
cuts the shock, and potentially introduces corner-terms that will
show up in the boundary stress energy as $\epsilon\to0$. See also
\citep{bonnor} for an earlier review of junction conditions. Instead
it is more straightforward to carry over the derivation of \citep{deHaro:2000vlm}
with the understanding that the metric components become generalized
functions in the time-direction.

In detail, we place a cutoff at $\rho=\epsilon$ in Fefferman-Graham
coordinates and study the metric in the interior of the gray-shaded
wedge in \ref{fig:Penrose-diagram-showing}. The null shell $u=-1$
becomes the surface $t=\log v$, $\rho=(1-\sqrt{v})^{2}/(1-v)$ (with
$v\in(0,1)$) while the shell at $v=1$ becomes the surface $t=-\log\left(-u\right)$,
$\rho=\left(1-\sqrt{-u}\right)^{2}/(1+u)$ with $u\in(-1,0)$. Outside
the wedge, the holographic stress tensor \citep{deHaro:2000vlm},
vanishes as $\epsilon\to0$.

Inside the wedge, Fefferman-Graham coordinates correspond to 
\[
u=-e^{-t/2}\frac{1-\rho}{1+\rho},\qquad v=e^{t/2}\frac{1-\rho}{1+\rho}
\]
where the metric is

\begin{align}
ds^{2} & =\frac{\ell^{2}}{\rho^{2}}\left(d\rho^{2}-\frac{1}{16}\left(\rho^{2}-1\right)^{2}dt^{2}+\frac{1}{\left(1+z\bar{z}\right)^{2}}\left|\left(1-\rho^{2}\right)\cosh\left(\frac{t}{2}\right)dz-\right.\right.\label{eq:fgmetric}\\
 & \left.\left.\left(1+z\bar{z}\right)^{2}\left(1+\rho^{2}+\left(\rho^{2}-1\right)\cosh\left(\frac{t}{2}\right)\right)d\bar{z}\bar{F}(\bar{z})\right|^{2}\right)\,.\nonumber 
\end{align}

The mapping between the CFT stress energy tensor and the Fefferman-Graham
expansion is \citep{deHaro:2000vlm}
\begin{equation}
\left\langle T_{ab}^{CFT}\right\rangle =\frac{3}{16\pi G_{N}}g_{ab}^{(3)}\,.\label{eq:holostress}
\end{equation}
To be able to apply \eqref{eq:holostress} we need to take into account
the step functions which localize the solution \eqref{eq:fgmetric}
to the interior of the shaded wedge, while outside we have the solution
with $\bar{F}=0$. To do this we Taylor expand the step functions
near $\rho=0$ that modulate the $\bar{F}$ terms, and the $F\bar{F}$
terms
\begin{align*}
\theta(t+4\tanh^{-1}\rho)-\theta(t-4\tanh^{-1}\rho)= & \left(\theta(t+4\tanh^{-1}\rho)-\theta(t-4\tanh^{-1}\rho)\right)^{2}\\
= & 8\rho\delta(t)+\frac{64}{3}\rho^{3}\delta''(t)+\frac{8}{3}\rho^{3}\delta(t)+\mathcal{O}(\rho^{4})\,.
\end{align*}
This $\rho$ dependence induces a $g_{ab}^{(3)}$ term in the metric,
and \eqref{eq:holostress} leads to
\begin{equation}
\left\langle T_{ab}^{CFT}\right\rangle =\frac{3}{16\pi G_{N}}g_{ab}^{(3)}=-\frac{2\ell^{2}}{\pi G_{N}}\left(\begin{array}{ccc}
0 & 0 & 0\\
0 & F(z) & 0\\
0 & 0 & \bar{F}(\bar{z})
\end{array}\right)\delta(t)\,.\label{eq:2dstress}
\end{equation}
The lower order components $g_{ab}^{(0)}$ and $g_{ab}^{(2)}$ simply
match the vacuum solution with $F(z)=0$.

This expectation value \ref{eq:2dstress} is exactly what one expects
for the vacuum expectation value of a two-dimensional defect CFT inserted
at $t=0$ after undergoing a holomorphic coordinate transformation
$z'=f(z)$
\begin{equation}
\left\langle T_{zz}^{2d,CFT}\right\rangle =-\frac{c}{12}\left\{ f(z),z\right\} \label{eq:onepoint}
\end{equation}
where $\left\{ f(z),z\right\} =F(z)$ is the Schwarzian derivative
\eqref{eq:schwarzian}. Recall this is indeed the coordinate transformation
on the boundary \eqref{eq:superrotate}. We therefore tentatively
identify the central charge of the 2d defect CFT with
\[
c=\frac{24\ell^{2}}{\pi G_{N}}\,.
\]
Clearly the flat space limit will then correspond to a large central
charge limit if $G_{N}$ is held fixed.

The one-point function \eqref{eq:onepoint} can be evaluated for a
general analytic diffeomorphism that maps the $t=0$ slice to itself.
One may therefore use the one-point function to iteratively generate
insertions of the boundary stress tensor $T_{zz}^{2d}$ on the $t=0$
slice. Therefore we expect to get a general set of $n-$point functions
of the 2d stress tensor that are constrained by the Virasoro algebra.

From the perspective of general 3d CFT this behavior is surprising.
We expect linear couplings between the 2d operators and the 3d operators,
and at best the 3d theory will enjoy a local $SO(3,2)$ symmetry,
not an infinite dimensional Virasoro symmetry. We therefore do not
expect the 2d theory to have an exact Virasoro symmetry constraining
its correlation functions. On the other hand we can only trust the
gravity results (of which \eqref{eq:onepoint} is an example) for
$\ell^{2}\gg G_{N}$ so we are necessarily working in a large central
charge limit. It is possible then for an enhanced symmetry to emerge,
broken by perturbative corrections in a $1/c$ expansion. One may
view the 3d CFT correlators (including mixed correlators involving
insertions at $t=0$) to be determined by usual holographic map from
gravity correlators to boundary correlators. It will be interesting
to work out the details of this map to see whether there is a sense
Virasoro symmetry constrains the correlators of general operators
at $t=0$, but we leave this for future work.

In summary, we conjecture the expectation value \eqref{eq:2dstress}
is most simply explained by the insertion of a boundary two-dimensional
conformal field theory at the surface $t=0$ within the three-dimensional
$dS_{3}$ boundary. Recall that the shock wave solutions we are studying
are constructed by gluing together two patches with the transformation
$z\rightarrow f(z)$ along the shock. The expectation values of the
stress tensor on the plane vanish in the three-dimensional vacuum
state ($f(z)=z)$, and we are left with the Schwarzian term above
when $f(z)$ is chosen more generally. This is certainly a highly
non-trivial proposal and many more consistency checks should be made
to ascertain its validity. 

\section{Asymptotically Flat Limit}

To construct the asymptotically flat limit, we take the metric in
the form \eqref{eq:implode_explode} and define rescaled coordinates
$\tilde{u}=\ell u,\,\tilde{v}=\ell v$. Taking the limit $\ell\to\infty$
with $\tilde{u},\tilde{v}$ fixed, we get
\[
ds^{2}=-4d\tilde{u}d\tilde{v}+\frac{4}{(1+z\bar{z})^{2}}\left|(\tilde{u}-\tilde{v})dz+\tilde{\bar{F}}(\bar{z})d\bar{z}(1+z\bar{z})^{2}\right|^{2}
\]
where $\bar{\tilde{F}}=\ell\bar{F}$. Next we transform to Bondi coordinates
on $\mathcal{I}^{+}$ defining
\[
\tilde{v}=\tilde{u}+\sqrt{\xi^{2}+\left(1+z\bar{z}\right)^{4}\left|\tilde{F}(z)\right|^{2}}
\]
and replacing $\tilde{u}\to u/2$ and $\tilde{F}\to F$ to give
\begin{align*}
ds^{2} & =-du^{2}-\left(2-\frac{1}{\xi^{2}}(1+|z|^{2})^{4}\left|F(z)\right|^{2}\right)dud\xi-\frac{1}{\xi}\partial_{z}\left(\left(1+|z|^{2}\right)^{4}\left|F(z)\right|^{2}\right)dzdu-\\
 & \frac{1}{\xi}\partial_{\bar{z}}\left(\left(1+|z|^{2}\right)^{4}\left|F(z)\right|^{2}\right)d\bar{z}du-4\xi F(z)dz^{2}-4\xi\bar{F}(\bar{z})d\bar{z}^{2}+\\
 & \xi^{2}\frac{4}{\left(1+|z|^{2}\right)^{2}}dzd\bar{z}+8(1+|z|^{2})^{2}\left|F(z)\right|^{2}dzd\bar{z}+\cdots
\end{align*}
where $\cdots$ refers to higher order terms in the $1/\xi$ expansion.
Reading off the Bondi parameters following the notation of \citep{Barnich:2011mi}
we find
\[
M=0,\qquad N_{AB}=0,\qquad C_{AB}=\left(\begin{array}{cc}
-4F(z) & 0\\
0 & -4\bar{F}(\bar{z})
\end{array}\right),\qquad D_{AB}=0,\qquad N_{A}=-\frac{3}{32}\partial_{A}C_{CD}C^{CD}
\]
here the indices $A,B$ run over $z,\bar{z}$. Putting these together,
we compute the integrable component of the superrotation charge \citep{Barnich:2011mi}
as
\begin{align*}
Q_{Y} & =\frac{1}{16\pi G_{N}}\int d^{2}z\frac{2}{(1+|z|^{2})^{2}}Y^{A}\left(2N_{A}+\frac{1}{16}\partial_{A}C_{CD}C^{CD}\right)\\
 & =\frac{1}{16\pi G_{N}}\int d^{2}z\frac{2}{(1+|z|^{2})^{2}}Y^{A}\left(-\frac{1}{8}\partial_{A}C_{CD}C^{CD}\right)
\end{align*}
where $Y^{A}(z,\bar{z})$ is a vector field on the 2-sphere that can
be chosen to compute the desired moment of the integrand. The superrotation
charge is therefore in general non-zero at quadratic order in the
perturbation.\footnote{There are appear to be various expressions for the superrotation charges
in the literature; see e.g. \textcolor{red}{\citep{Hawking:2016sgy}.}
The different choices lead to different values of the charges we are
evaluating here, so we have simply opted to go with the definitions
of \citep{Barnich:2011mi}.}

These solutions with vanishing ADM momentum and non-vanishing $C_{AB}$
are noted in \citep{Hawking:2016sgy}. For example, supertranslations
can be used to induce non-vanishing $C_{AB}$ from the state where
all Bondi parameters vanish. As in that case, there exist transformed
Poincare generators that leave the solution invariant, as is clear
from the construction of the solution as a diffeomorphism of flat
spacetime \citep{Hogan:1995qb}. These solutions are related to the
results of \citep{Chen:2014uma} where solutions with vanishing ADM
momentum but non-vanishing ADM angular momentum are studied (though
in the present work $F(z)=0$ for any globally defined $SL(2,C)$
transformation, so we only expect to see superangular momenta).

We note a linearized version of the Schwarzian term appears in the
work of \citep{Kapec:2014opa} when $N_{AB}$ and $C_{AB}$ undergo
a superrotation. In the present work the Bondi news $N_{AB}$ vanishes
except for a Dirac delta function term localized on the shock. In
the limit considered in this section the shock itself does not appear
in the asymptotically flat region. However shocks that cross the asymptotically
flat region can similarly be constructed, and are briefly commented
on below. 

\section{Discussion}

In the present work, we have constructed a shock solution in asymptotically
local anti-de Sitter spacetime and found the corresponding boundary
stress tensor has an interpretation in terms of a two-dimensional
Euclidean conformal field theory living on a timeslice within a three-dimensional
conformal field theory. In the limit of asymptotically flat spacetime
the shock gives a patch of spacetime with nontrivial superrotation
charges. In a more general context, one may insert 2d CFT operators
and build more general Virasoro charges along with the holomorphic
coordinate transformation that appears in the shock solution. One
may follow through the usual construction of perturbative bulk fields
around this solution \citep{Hamilton:2005ju,Hamilton:2006az} and
there will be contributions from the 2d CFT and the more standard
contributions from the 3d CFT. The contributions from operators in
the 2d CFT will contain the leading order information about the BMS
charges, while the 3d CFT will only have the usual global conformal
charges $SO(3,2)$ after explicitly breaking the symmetry with the
defect. These global charges contract to the 4d Poincare group in
the flat space limit $ISO(3,1)$ \citep{Compere:2019bua}.

One might also consider shocks crossing the asymptotically flat region
that can give rise to other defect timeslices in the 3d CFT. It is
possible there is some connection between this picture and the suggestion
that strings pierce $\mathcal{I}^{+}$ in the asymptotically flat
limit, destroying asymptotic flatness in general \citep{Penrose:1972xrn}.
Conversely, one might rule out such additional shocks if one insists
on a well-defined asymptotically flat limit. Note there do exist Penrose
shock solutions involving more general BMS transformations \citep{OLoughlin:2018ebk,Strominger:2016wns}
and it would be interesting to study these from the present perspective. 

A primary motivation for the present work is to gain better insight
into how the BMS symmetries generalize to the quantum case. The proposal
of the present work is that the generalization of a class of spacetimes
with nontrivial superrotation charges to asymptotically AdS spaces
are dual to a conventional 3d CFT with a 2d CFT living at a particular
timeslice. Such a system should have a self-contained quantum description.
Another approach to accommodate BMS like symmetries in AdS is to allow
for more general boundary conditions \citep{Compere:2019bua,Compere:2020lrt},
however this avenue appears to lead to boundary gravity theories that
seem difficult to describe.

The idea of a defect CFT living at a particular timeslice is perhaps
a rather novel idea, however in Euclidean signature this is a relatively
familiar construction. In Lorentzian signature, one can view the defect
as the coincident limit of a pair of local operator quenches. This
coincident limit takes one out of the realm of the 3d CFT (or equivalently
involves infinite energies in the 3d CFT) and requires the specification
of the 2d CFT to make it well-defined. It will be very interesting
to further study the details of this construction to better develop
the structure of this 2d CFT and its coupling to the 3d CFT. Thus
far, we have only been able to read off the central charge and see
the Schwarzian vacuum energy emerge.\vfill{}

\begin{acknowledgments}
D.L. is supported in part by DOE grant de-sc0010010. D.L. thanks R.
Hingley and K. Prabhu for helpful discussions.
\end{acknowledgments}

\bibliographystyle{utphys}
\bibliography{bms-ads}

\end{document}